\documentclass[twocolumn,showpacs,superscriptaddress,preprintnumbers,amsmath,amssymb]{revtex4}

\usepackage{graphicx}
\usepackage{dcolumn}
\usepackage{bm}

\newcommand{\bq}{\begin{equation}}
\newcommand{\eq}{\end{equation}}
\newcommand{\bqa}{\begin{eqnarray}}
\newcommand{\eqa}{\end{eqnarray}}
\newcommand{\nn}{\nonumber \\}

\def\be     {\begin{equation}}
\def\ee     {\end{equation}}
\def\bea        {\begin{eqnarray}}
\def\eea        {\end{eqnarray}}
\def\bnn    {\begin{eqnarray*}}
\def\enn    {\end{eqnarray*}}

\begin{document}

\title{$z = 3$ antiferromagnetic quantum critical point :
U(1) slave-fermion theory of Anderson lattice model}
\author{Ki-Seok Kim}

\affiliation{Asia Pacific Center for Theoretical Physics, Hogil
Kim Memorial building 5th floor, POSTECH, Hyoja-dong, Namgu,
Pohang 790-784, Korea}

\author{Chenglong Jia}

\affiliation{Institut f\"ur Physik, Martin-Luther Universit\"at
Halle-Wittenberg, 06120 Halle (Saale), Germany }



\date{\today}

\begin{abstract}
We find the dynamical exponent $z = 3$ antiferromagnetic (AF)
quantum critical point (QCP) in the heavy fermion quantum
transition beyond the standard framework of the
Hertz-Moriya-Millis theory with $z = 2$. Based on the U(1)
slave-fermion representation of the Anderson lattice model, we
show the continuous transition from an antiferromagnetic metal to
a heavy fermion Fermi liquid, where the heavy fermion phase
consists of two fluids, differentiated from the slave-boson
theory. Thermodynamics and transport of the $z = 3$ AF QCP are
shown to be consistent with the well known non-Fermi liquid
physics such as the divergent Gr\"{u}neisen ratio with an exponent
$2/3$ and temperature-linear resistivity. In particular, the
uniform spin susceptibility turns out to diverge with an exponent
$2/3$, the hallmark of the $z = 3$ AF QCP described by deconfined
bosonic spinons.
\end{abstract}


\maketitle

The continuous quantum transition from an antiferromagnetic (AF)
metal to a heavy fermion (HF) Fermi liquid has been one of the
central interests in strongly correlated electrons since its
quantum critical point (QCP) is beyond the description of the
Landau's Fermi liquid theory and Landau-Ginzburg-Wilson's
framework for phase transitions, two cornerstones in modern theory
of metals \cite{HFQCP}. Thermodynamics such as the divergent
Gr\"{u}neisen ratio with an exponent $2/3$ \cite{GR} and non-Fermi
liquid transport of temperature-linear resistivity \cite{TR} are
difficult to describe in the standard framework based on the weak
coupling approach, that is, Hertz-Moriya-Millis (HMM) theory for
the AF transition \cite{HMM}.

Recently, critical hybridization fluctuations described by the
dynamical exponent $z = 3$ were proposed in the U(1) slave-boson
description for the HF transition \cite{Paul,Pepin}, explaining
both the thermodynamics \cite{Kim_GR} and transport \cite{Kim_TR}
qualitatively well. However, quantum fluctuations of spin dynamics
are overestimated in the slave-boson representation, giving rise
to the so called fractionalized Fermi liquid \cite{VSS_KB} instead
of an AF metal. In addition, introduction of the hybridization
order parameter allows an artificial finite temperature
transition, not observed but identified with the crossover to the
HF phase in experiments.

In this paper we revisit the AF to HF quantum transition, based on
the U(1) slave-fermion representation incorporating AF
correlations well. On the contrary to the common wisdom, we show
that the slave-fermion description allows the HF liquid, resorting
to the Luttinger theorem \cite{Luttinger_Theorem} and an explicit
mean-field analysis. In particular, the artificial HF transition
at finite temperatures does not appear in the slave-fermion
representation, where the HF phase is not described by band
hybridization, different from the slave-boson theory.

Our main result is that critical spin dynamics described by
bosonic spinons is governed by the dynamical exponent $z = 3$
resulting from Landau damping of conduction electrons and
fermionic holons, basically the same as critical hybridization
fluctuations in the slave-boson theory \cite{Paul,Pepin} but
completely different from the HMM theory with $z = 2$. As a
result, the slave-fermion theory reproduces the anomalous
thermodynamics \cite{Kim_GR} and non-Fermi liquid transport
\cite{Kim_TR} of the slave-boson description.

The hallmark of the $z = 3$ AF QCP is seen from the uniform spin
susceptibility diverging with an exponent $2/3$, consistent with
an experiment $0.72 \pm 0.05$ \cite{INS}. This result turns out to
originate from the bosonic nature of fractionalized spinon
excitations. Scaling of several quantities is shown for various
scenarios in Table I.


\begin{table}[ht]
\begin{tabular}{ccccc}
\; & $z$ \& $\nu$ \; & $\Gamma (T)$ \; & $\chi (T)$ \; & $\rho
(T)$ \; \nn \hline SF QCP \; & 3 \& 1/2 \; \; & $T^{-2/3}$ \; \; &
$T^{-2/3}$ \; \; & $T \ln (2T/E^{*})$ \; \;  \nn \hline KB QCP \;
& 3 \& 1/2 \; \; & $T^{-2/3}$ \; \; & const. \; \; & $T \ln
(2T/E^{*})$ \; \; \nn \hline Local QCP \; & - $\infty$ \& ?? \; \;
& $T^{-0.7}$ \; \; & $T^{-0.75}$ \; \; & ?? \; \; \nn \hline HMM
QCP \; & 2 \& 1/2 \; \; & $T^{-1}$ \; \; & const. \; \; &
$T^{3/2}$ \; \; \nn \hline
\end{tabular}
\caption{Scaling of Gr\"{u}neisen ratio $\Gamma (T)$, uniform spin
susceptibility $\chi (T)$, and resistivity $\rho (T)$ with
dynamical $z$ and correlation-length $\nu$ exponents in $d=3$ for
the slave-fermion (SF), Kondo breakdown (KB), Local, and
Hertz-Moriya-Millis (HMM) QCP scenarios, respectively.}
\end{table}


We point out that our present study generalizes the slave-fermion
description for ferromagnetism in the Anderson lattice model
\cite{SF_FM} into the case of antiferromagnetism and the
slave-fermion study for the two impurity problem \cite{SF_NCA}
into the case of an impurity lattice, respectively. The $z = 3$ AF
QCP is consistent with the previous slave-fermion study
\cite{Pepin_SF}, where both the existence of the QCP and dynamics
of holons
are assumed, but not fully justified.

We start from the U(1) slave-fermion representation of an
effective Anderson lattice model \bqa && Z =
\int{Dc_{in\sigma}Db_{in\sigma}Df_{i}D\Delta_{ij}D\chi_{ij}^{b}D\chi_{ij}^{f}D\lambda_{i}}
e^{- \int_{0}^{\beta}{d\tau} L} , \nn && L = L_{c} + L_{f} + L_{b}
+ L_{V} + L_{0} , \nn && L_{c} =
\sum_{i}c_{in\sigma}^{\dagger}(\partial_{\tau} - \mu)c_{in\sigma}
- t\sum_{\langle ij\rangle}(c_{in\sigma}^{\dagger}c_{jn\sigma} +
H.c.) , \nn && L_{f} = \sum_{i} f_{i}^{\dagger}(\partial_{\tau} +
i\lambda_{i})f_{i} + \alpha t\sum_{\langle ij\rangle}
(f_{j}^{\dagger}\chi_{ij}^{b*}f_{i} + H.c.) , \nn && L_{b} =
\sum_{i} b_{in\sigma}^{\dagger}(\partial_{\tau} + \epsilon_{f} +
i\lambda_{i})b_{in\sigma} - \alpha t\sum_{\langle ij\rangle} (
b_{in\sigma}^{\dagger}\chi_{ij}^{f}b_{jn\sigma} \nn && + H.c.) - J
\sum_{\langle ij\rangle} (
\Delta_{ij}^{*}\epsilon_{\sigma\sigma'}b_{in\sigma}b_{jn\sigma'} +
H.c. ) , \nn && L_{V} = V \sum_{i}
(c_{in\sigma}^{\dagger}b_{in\sigma}f_{i}^{\dagger} + H.c.) , \nn
&& L_{0} = \alpha t\sum_{\langle ij\rangle}
(\chi_{ij}^{b*}\chi_{ij}^{f} + H.c.) + N J \sum_{\langle
ij\rangle} |\Delta_{ij}|^{2} \nn && - i \sum_{i} 2 N S\lambda_{i}
, \eqa
where the hybridization term $V$ competes with the AF correlation
term $J$ for localized electrons, modelled as the nearest neighbor
spin-exchange interaction. $L_{c}$ describes dynamics of
conduction electrons $c_{in\sigma}$, where $\mu$ and $t$ are their
chemical potential and kinetic energy, respectively. $L_{f}$ and
$L_{b}$ govern dynamics of localized electrons, decomposed with
fermionic holons $f_{i}$ and bosonic spinons $b_{in\sigma}$, where
local AF correlations $\Delta_{ij}$ are introduced in the Sp(N)
representation for the spin-exchange term $J$ with an index $n =
1, ..., N$ \cite{Sachdev_SpN} and an almost flat band with $\alpha
\ll 1$ is allowed \cite{Pepin} to describe hopping of holons
$\chi_{ij}^{b}$ and spinons $\chi_{ij}^{f}$, respectively.
$\epsilon_{f}$ is an energy level for the flat band, and
$\lambda_{i}$ is a Lagrange multiplier field to impose the
slave-fermion constraint. $L_{V}$ is the hybridization term,
involving conduction electrons, holons, and spinons. $L_{0}$
represents condensation energy with $N = 1$ and $S = 1/2$ in the
physical case.

In the limit of $V \rightarrow 0$ the slave-fermion Lagrangian is
reduced to two decoupled sectors for conduction electrons and
spinons, where ferromagnetic (FM) correlations $\chi_{ij}^{f}$
vanish in the spinon sector, recovering the Schwinger-boson theory
for the half filled quantum antiferromagnet
\cite{Schwinger_Boson}. In this respect the present problem
generalizes the Schwinger-boson theory, turning on hybridization
fluctuations to cause "hole doping" in the localized band,
represented by fermionic holons. Particulary, hybridization
fluctuations give rise to FM correlations, weakening AF
correlations $\Delta_{ij}$ and destroying the AF order $\langle
b_{in\sigma} \rangle = 0$.

The resulting paramagnetic phase turns out to be a HF metal,
differentiated from that of the slave-boson theory described by
band hybridization, where the localized band is decoupled with the
conduction band due to gapping of spinons $\langle b_{in\sigma}
\rangle = 0$. Instead, the effective chemical potential denoted by
$i\lambda_{i}$ is changed by the hybridization coupling constant
$V$, filling holons to the almost flat band. In this respect the
HF phase of the slave-fermion representation consists of two kinds
of fluids, corresponding to light fermions of conduction electrons
and heavy fermions of holons, respectively.

The presence of the HF phase in the slave-fermion description can
be argued from the Luttinger theorem. Inserting the total electric
charge $\delta = c_{in\sigma}^{\dagger}c_{in\sigma} -
f_{i}^{\dagger}f_{i}$ into the single occupancy constraint
$b_{in\sigma}^{\dagger} b_{in\sigma} + f_{i}^{\dagger}f_{i} + 2N
\Delta_{i}^{\dagger}\Delta_{i} = 2NS$ modified by the presence of
local singlets $\Delta_{i}$, we find
$c_{in\sigma}^{\dagger}c_{in\sigma} = 2NS + \delta - 2N
\Delta_{i}^{\dagger}\Delta_{i} - b_{in\sigma}^{\dagger}
b_{in\sigma}$. As a result, the Luttinger theorem holds, given by
$\frac{V_{FS}^{el}}{2\pi} = 1 + \delta - 2 \Delta^{2}$ and
implying the large Fermi surface, where $V_{FS}^{el}$ is the
volume of the electron Fermi surface and $\delta$ is the density
of conduction electrons in the decoupling limit. It is interesting
to observe that the area of the Fermi surface is not $1 + \delta$
but smaller owing to the presence of AF correlations.

\begin{figure}[t]
\vspace{4cm} \includegraphics{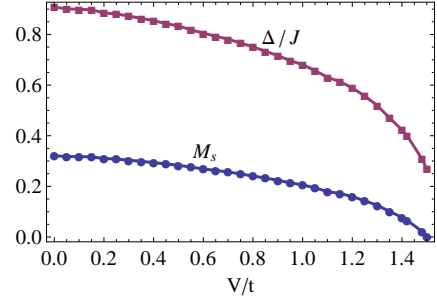} \caption{ (Color online)
Hybridization-fluctuation-induced AF QCP, where condensation
probability $M_{s} \propto |\langle b_{n\sigma} \rangle|^{2}$
(blue) vanishes but antiferromagnetic correlations $\Delta$
(purple) still exist.} \label{fig1} \vspace{-0.5cm}
\end{figure}

Performing the mean-field approximation of uniform hopping
$\chi_{ij}^{f(b)} \rightarrow \chi_{f(b)}$, pairing $\Delta_{ij}
\rightarrow \Delta$, and chemical potential $i\lambda_{i}
\rightarrow \lambda$, we find the slave-fermion mean-field phase
diagram for the Anderson lattice model (Fig. 1). Actually, we see
that the spinon condensation amplitude vanishes at the critical
hybridization strength $V_{c}$. In the AF phase ($V < V_{c}$) band
hybridization is allowed, but the area of the Fermi surface will
be small, proportional to $\delta$ because the effective chemical
potential of holons is almost on the top of the holon band and the
density of holons is vanishingly small. Enhancing the
hybridization coupling constant, the holon chemical potential
shifts to the lower part, filling holons into the flat band and
causing heavy fermions. In this description the HF transition at
finite temperatures turns into crossover, where the crossover
temperature $T_{FL}$ is given by gap of spinon excitations $T_{FL}
\sim \xi_{s}^{-1}$ with the correlation length $\xi_{s} =
[(\lambda - 2 d \alpha t \chi_{f})^{2} - (2 d \Delta
)^{2}]^{-1/2}$ since scattering of conduction electrons and holons
with spinon fluctuations is suppressed below this temperature
allowing Fermi liquid physics.

Thermodynamics around the slave-fermion AF QCP can be understood
from the scaling expression of the free energy \bqa && f_{s}(r,T)
= b^{-(d+z)} f_{r}(r b^{1/\nu}, T b^{z}) , \eqa where $r \propto
|V - V_{c}|$ is an external parameter associated with mass for
critical fluctuations and $b$ is a scaling parameter with
dimension of length. The dynamical exponent $z$ tells the nature
of critical fluctuations, i.e., their dispersion relation $\Omega
\propto q^{z}$, and the correlation-length exponent $\nu$ gives
how the correlation length $\xi$ changes with respect to the
external parameter $r$, i.e., $\xi \propto |r|^{-\nu}$. Our main
problem is to derive this scaling free energy from the
slave-fermion theory. Actually, this was performed in the
slave-boson context, constructing the Luttinger-Ward (LW)
functional in the Eliashberg approximation \cite{Kim_LW}, where
momentum dependence in fermion self-energies and vertex
corrections are neglected, allowing us to introduce one loop-level
quantum corrections fully self-consistently.
It was explicitly demonstrated that the Eliashberg framework is
"exact" in the large $N$ limit \cite{FMQCP}.

Our main discovery is that dynamics of spinon fluctuations is
described by $z = 3$ critical theory due to Landau damping of
electron-holon polarization above an intrinsic energy scale
$E^{*}$, while by $z = 1$ O(4) nonlinear $\sigma$ model
\cite{Sachdev_O4} below $E^{*}$. The energy scale $E^{*} \propto
\alpha D (q^{*}/k_{F}^{c})^{3}$ originates from the mismatch
$q^{*} = |k_{F}^{f} - k_{F}^{c}|$ of the Fermi surfaces of the
conduction electrons $k_{F}^{c}$ and holons $k_{F}^{f}$, shown to
vary from ${\cal O}(10^{0})$ $mK$ to ${\cal O}(10^{2})$ $mK$
\cite{Paul,Pepin}. Actually, inserting the Landau damping
self-energy $\Pi_{b}(q,i\Omega) = \gamma_{b} \frac{|\Omega|}{q}$
with the damping coefficient $\gamma_{b} = \frac{\pi}{2}
\frac{V^{2}\rho_{c}}{v_{F}^{f}}$ into the spinon's full
propagator, where $\rho_{c}$ is the density of states for
conduction electrons and $v_{F}^{f}$ is the holon velocity, we
find their $z = 3$ dynamics \bqa && \Im D_{b}(q,\Omega) \approx -
\frac{\gamma}{2\gamma_{b}} \frac{ \gamma \Omega q }{ q^{6} +
\gamma^{2} \Omega^{2} } \eqa with $\gamma \equiv (2\gamma_{b})(2 d
\Delta/v_{s}^{2})$, where $v_{s} = \sqrt{2 [\alpha t \chi_{f}
(\lambda - 2 d \alpha t \chi_{f}) + (2 d \Delta^{2} )]}$ is the
velocity of spinons. Then, the correlation-length exponent is
given by the usual mean-field value $\nu = 1/2$ since the critical
theory is above its upper critical dimension in $d = 3$.

Inserting Eq. (3) into the LW expression of the free energy, we
find Eq. (2) with $z = 3$ and $\nu = 1/2$ for the singular part
above $E^{*}$ \cite{Kim_LW}. As a result, we obtain both the
logarithmic divergent specific heat coefficient and power-law
diverging thermal expansion coefficient, giving rise to the
divergent Gr\"{u}neisen ratio $T^{-1/\nu z}$ \cite{GR_Scaling}
with an exponent $2/3$ up to the logarithmic correction
\cite{Kim_GR}.


An important result of the $z = 3$ slave-fermion QCP is that the
dynamic uniform spin susceptibility diverges with an anomalous
exponent $2/3$. The transverse spin susceptibility is given by sum
of both spinon and electron susceptibilities \bqa
\chi_{t}^{+-}(q,\Omega) = \frac{N}{J(q)} \frac{1}{1 -
\frac{J(q)}{2N} \chi_{0b}^{+-}(q,\Omega)} +
\chi_{0c}^{+-}(q,\Omega) ,  \eqa where the spinon response is the
standard RPA expression with the momentum dependent exchange
coupling $J(q)$ and $\chi_{0b,c}^{+-}(q,i\Omega) \equiv -
\Bigl\langle S_{b,c}^{+}(q,i\Omega)S_{b,c}^{-}(-q,-i\Omega)
\Bigr\rangle_{c}$ are bare spin susceptibilities for spinons and
electrons, respectively, with the subscript $c$ meaning
"connected".

We find the spinon susceptibility \bqa && \Im
\chi^{+-}_{0b}(\Omega) \approx \frac{\mathcal{C}_{s}
N\gamma^{5/3}}{8\pi^{4}\gamma_{b}^{2}} \Omega^{2/3} , \eqa where
$\mathcal{C}_{s} = \int_{-1}^{0} d y \int^{\infty}_{0} d x x^{2}
\frac{ x y }{ x^{6} + y^{2} } \frac{ x( y + 1) }{ x^{6} + ( y +
1)^{2} } \approx 0.52$. One can check that this expression
coincides with the scaling theory, meaning that replacement of $r$
with the uniform magnetic field $h$ in Eq. (2) gives rise to
$\chi_{s}^{b}(h,T) = - T^{ (d+z)/z -2/z\nu} \frac{\partial^{2}
f_{r}^{b}(x, 1)}{\partial x^{2}}\Bigr|_{x = h T^{-1/z\nu}} \propto
T^{2/3}$ in $d = 3$. On the other hand, the electron spin
susceptibility is $\chi^{+-}_{0c}(q,\Omega) = \chi_{c}^{p} +
\gamma_{c} \frac{|\Omega|}{q}$ in $|\Omega| \ll q$, where
$\chi_{c}^{p}$ is the Pauli susceptibility and the damping
coefficient is $\gamma_{c} = \frac{N\rho_{c}}{v_{F}^{c}}$. As a
result, critical spinon excitations contribute to the spin
susceptibility dominantly, given by (Top in Fig. 2) \bqa && \Im
\chi_{t}^{+-}(\Omega) \approx \frac{2N^{2}}{(dJ_{c})^{2}} [\Im
\chi_{0b}^{+-}(\Omega)]^{-1} \propto \Omega^{-2/3} , \nonumber
\eqa where $J_{c}$ is the critical value for the slave-fermion
QCP.

\begin{figure}[t]
\vspace{7cm} \includegraphics{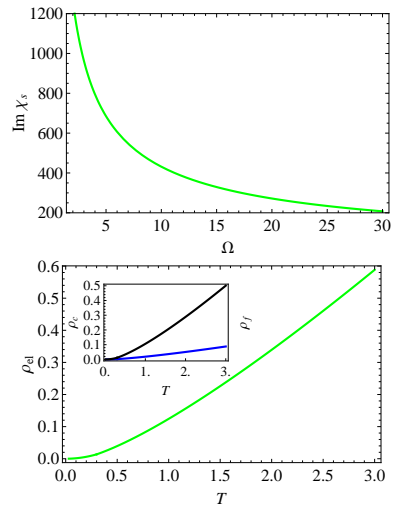} \caption{ (Color online) Top:
Uniform dynamic spin susceptibility described by deconfined
bosonic spinons scales with $\Omega^{-2/3}$ in the $z = 3$ regime.
Bottom: Temperature linear electrical resistivity. Right-inset:
Resistivity from conduction electrons (Blue) and that from holons
(Black).} \label{fig2} \vspace{-0.5cm}
\end{figure}



The diverging uniform spin susceptibility with the exponent $2/3$
is identified with the hallmark of the $z = 3$ AF QCP described by
"deconfined" bosonic spinons. If the QCP falls into the
universality class of the HMM theory, the exponent for divergence
is half of the present value \cite{HMM_1_3}. Strength of
divergence is enhanced by fractionalization, giving rise to
multiple correlations of deconfined degrees of freedom
\cite{Senthil_Deconfinement}. If spinon excitations were fermionic
in nature, the uniform susceptibility would be governed by the
Pauli susceptibility. In addition, the singular behavior of the
uniform susceptibility is a special feature of the slave-fermion
description because it is difficult to find in the HMM framework,
although it allows the Curie-Weiss behavior at the AF wave vector
in itinerant electrons \cite{Takimoto}.

The $z = 3$ AF QCP results in the temperature linear resistivity.
Electrical conductivity is given by the Ioffe-Larkin composition
rule due to the single occupancy constraint \cite{Ioffe_Larkin}
\bqa && \sigma_{el} = \sigma_{c} + \frac{\sigma_{f}(\sigma_{b} +
\sigma_{\Delta})}{\sigma_{f} + \sigma_{b} + \sigma_{\Delta}}
\approx \sigma_{c} + \sigma_{f} , \eqa where the subscript
represents each field and the last expression resorts to
$\sigma_{\Delta} \rightarrow \infty$. This result is reasonable
because only conduction electrons and holons carry electric
charges.


Using the Kubo formula expressed by the current-current
correlation function, we obtain \bqa \sigma_{c}(T) \approx
\frac{2\mathcal{C} N}{\pi} \frac{\rho_{c}
v_{F}^{c2}}{\Im\Sigma_{c}(k_{F}^{c},T)} , ~~ \sigma_{f}(T) \approx
\frac{ \mathcal{C} }{\pi} \frac{\rho_{f}
v_{F}^{f2}}{\Im\Sigma_{f}(k_{F}^{f},T)}    \eqa in the one loop
approximation with $\mathcal{C} = \int_{-\infty}^{\infty}{d y}
\frac{1}{(y^{2}+1)^{2}} = \frac{\pi}{2}$, where
$\Im\Sigma_{c,f}(k_{F}^{c,f},T)$ are imaginary parts of the
self-energies \cite{Kim_TR}. Scattering of fermions with $z = 3$
critical fluctuations results in \bqa && \Im \Sigma_{c}(T>E^{*}) =
\frac{\gamma }{2\gamma_{b}} \frac{ V^{2}}{12\pi^{2} v_{F}^{f}} T
\ln \Bigl( \frac{2T}{E^{*}} \Bigr) , \nn && \Im
\Sigma_{c}(T<E^{*}) = \frac{\gamma }{2\gamma_{b}} \frac{
V^{2}}{12\pi^{2} v_{F}^{f}} \frac{T^{2}}{E^{*}} \ln 2 \eqa for
conduction electrons and \bqa && \Im \Sigma_{f}(T>E^{*}) = 2N
\frac{\gamma }{2\gamma_{b}} \frac{ V^{2}}{12\pi^{2} v_{F}^{c}} T
\ln \Bigl( \frac{2T}{E^{*}} \Bigr) , \nn && \Im
\Sigma_{f}(T<E^{*}) = 2N \frac{\gamma }{2\gamma_{b}} \frac{
V^{2}}{12\pi^{2} v_{F}^{c}} \frac{T^{2}}{E^{*}} \ln 2 \eqa for
holons, basically the same as those of the slave-boson theory
\cite{Kim_TR}.

Bottom in Fig. 2 shows the quasi-linear behavior in temperature
for electrical resistivity above $E^{*}$, resulting from the
dominant $z = 3$ scattering with spinon fluctuations. The
$T$-linear relaxation time in transport is typical of the scaling
of the free energy with $z = 3$ and $\nu = 1/2$, provided a
mechanism for decaying the current is present in the theory
\cite{Kim_TR}.

In the present study gauge fluctuations associated with spin
collective modes are not taken into account, particulary, in the
heavy fermion phase. Actually, this consideration is protected in
the low temperature regime, where gauge fluctuations are gapped.
The U(1) gauge symmetry is reduced to Z$_{2}$ owing to the
presence of both AF ($\Delta$) and FM ($\chi_{f}$) correlations.
Introducing two energy scales of $T_{\Delta} \propto J \Delta^{2}$
for AF correlations and $T_{\chi} \propto \alpha t \chi_{b}
\chi_{f}$ for FM fluctuations, $T_{\Delta} \gg T_{\chi}$ is
expected because it is indeed true in the decoupling limit and
$\alpha \ll 1$ preserves this expectation. In the regime of $T
\leq T_{\chi}$ Z$_{2}$ gauge fluctuations are gapped, justifying
our treatment, where our rough estimate tells $T_{\chi} \sim
\mathcal{O}(10^{1}) K$ larger than $E^{*}$. Even in $T \geq
T_{\chi}$, dynamics of gauge fluctuations will be described by $z
= 3$ due to Landau damping of holon excitations, which cannot
change our present results qualitatively, but modifying those
quantitatively \cite{Kim_GR,Kim_TR}.

In this paper we find $z = 3$ AF QCP in the slave-fermion
description for the HF quantum transition, where the HF phase is
composed of both light conduction electrons and heavy holons. The
$z = 3$ quantum criticality turns out to result from
fractionalization of spin excitations, giving rise to
the well known anomalous thermodynamics \cite{GR} and non-Fermi
liquid transport \cite{TR}. In particular, the bosonic nature of
deconfined spinons was shown to cause the divergent uniform spin
susceptibility with an anomalous scaling exponent $2/3$, the
hallmark of the present theory, qualitatively consistent with an
experiment \cite{INS}.


K.-S. Kim thanks C. P\'epin for introducing this problem and
helpful discussions on the initial stage. Fruitful discussions
with T. Takimoto are appreciated.

\end{document}